\documentclass[%
 reprint,
 amsmath,amssymb,
 aps,
]{revtex4-2}

\usepackage{tikz}
\usepackage{graphicx}
\usepackage{dcolumn}
\usepackage{subcaption}
\usepackage{bm}
\usepackage{float}

\begin{document}

\preprint{APS/123-QED}

\title{DFT and Tight-binding Insights into Inorganic Halide Perovskites}
\title{Wannierized Tight-binding Insights into Inorganic Halide Perovskites}
\title{Density Functional Leveraged Tight-binding Insights into Inorganic Halide Perovskites}
\author{Misbah Shaheen and Sheharyar Pervez}
 \email{sheharyar@giki.edu.pk}
\affiliation{%
    Faculty of Engineering Sciences\\
    Ghulam Ishaq Khan Institute of Engineering Sciences and Technology
}%

\date{\today}

\newcommand\Briii{CsPbBr$_3$}
\newcommand\BriiI{CsPbBr$_2$I}
\newcommand\BrIii{CsPbBrI$_2$}
\newcommand\Iiii{CsPbI$_3$}
\newcommand\px{$p_{x}$}
\newcommand\py{$p_{y}$}
\newcommand\pz{$p_{z}$}

\begin{abstract}

Compared to first principles studies like DFT (physically accurate, computationally expensive), the TB approach allows the disentanglement of the region of interest, to wit the fermi level. This allows the creation of simplified, highly interpretable models that give chemically grounded insights. In this paper we employ a Wannierized TB approach to tune the bandgap and electronic structure via anion exchange and investigate its impact on the orbital interactives of inorganic lead halide perovskites. Further adjustment of the hopping norm and maximum distance leads to a reduced TB model which regenerates the broad features of the band structure with a fraction of the parameters, thereby making the model simpler and highly interpretable.
We observe that due to the asymmetry of the unit cell and the delocalized nature of its orbitals, \BrIii\ follows a many-small hopping scheme different in character from the few-big strategy taken by the other perovskites.  
Finally we leverage the TB model to study the electronic and thermal transport properties of these materials. These insight enable the identification of optimal doping strategies that could enhance thermoelectric performance by improving carrier mobility or optimizing the Seebeck coefficient



\end{abstract}

\maketitle


\section{\label{introduction}Introduction}

Functional materials such as inorganic lead halide perovskites (ILHPs) are pivotal to several modern world technology for energy harvesting, energy conversion and waste heat recovery with which we may build a sustainable future \cite{ref1} \cite{IHPs_solarcell}. Their flexible structure coupled with interesting physico-chemical properties, such as strong light absorption, bandgap tunability, small exciton binding energy, good carrier mobility and Seebeck coefficient, and ultralow thermal conductivity across a range of compositions and phases; makes them promising candidates for photovoltaic and thermoelectric applications. \cite{https://doi.org/10.1002/advs.201903389}  

Both the crystal structure and the bandgap heavily dependent on the anion composition. 
Since the soft crystal structure of these materials facilitates diffusion of the halide ions through the crystal lattice, anion exchange driven dynamical disorder can be used to tune the bandgap, improve carrier transport, and consequently enhance thermal and optical profiles \cite{anion_exchange0}. The Goldschmidt tolerance factor $t$ and the octahedral factor ($\mu$), which measure geometric stability, are expected to lie between $0.8-1.1$ and $0.44-0.9$, respectively for a stable perovskite structure \cite{goldschmidt_gesetze_1926}. The exchange of the A-site and/or X-site restores the tolerance factor closer to 1, thus stabilizing the cubic structure at room temperature \cite{article} \cite{anionexchange}. 

To solve for the electronic structure, in order to predict the physical and chemical properties, we have at our disposal
first principles methods like DFT, which try to solve approximations of the many-body Schrodinger equation. Though more accurate, these are also computationally expensive.
Tight-binding models are a popular, computationally less intensive alternative where a good model Hamiltonian is one which captures key features of the region of interest with minimal requirements.

In this study we tune the cubic phase ILHPs bandgap via anion exchange and examine the resulting impact on electronic and thermoelectric properties using a synergetic density functional theory (DFT) and tight binding (TB) approach. The TB approach allows us to create a model that is chemically grounded and computationally simple. Further examination leads us to a reduced TB (redTB) model which is readily interpretable. The redTB model highlights the key elements of the ILHP Hamiltonian. Finally we categorize the materials according to a many-small vs. few-big (MSFB) scheme based on the distribution of hopping spectrum. 

The rest of the paper is organized as follows. First we present the background on DFT, the tight-binding method, and transport properties. We then detail our methodology and finally the interpretation of results where we present the reduced TB method and discuss the thermoelectric profiles.


\section{\label{preliminaries}Preliminaries}

\subsection{\label{dftandtb}Density Fucntional Theory (DFT) and Tight Binding (TB)}
Conventional Kohn-Sham DFT framework uses Bloch states $\psi_{ nk}(\mathbf{r})$ to solve the KS equations  for a fictitious system of non-interacting particles instead of the many body Schrodinger equation of the fully interacting Hamiltonian. The eigenstates extend over the Brillouin zone, and are required to be orthogonal, which makes the computational complexity grow cubically with system size. 

Subjecting the DFT deduced Bloch orbitals $\psi_{n,k}(r)$ to a unitary transformation matrix $\mathcal{U}$, one can generate localized orthonormal orbitals, called Wannier functions (WFs) $w_n(\mathbf{r}-\mathbf{R})$, identified by the band index $n$, and the unit cell vector $\mathbf{R}$ \cite{WFs}.

\begin{equation}
    |w_n(\mathbf{r}-\mathbf{R})\rangle= \frac{V}{(2\pi)^3} \int_{BZ} dk \sum_{m=1}^{N} e^{i.k.R}\mathcal{U}_{mn}(k) |\psi_{n,k}\rangle
    \label{WF}
\end{equation} 

where $\mathcal{U}_{mn}(k)$ are the elements of the unitary transformation matrix at a specific $k$. The degree of localization is determined by the spread functional $\Omega$ as:

\begin{equation}
    \Omega = \sum_{n=1}^{N} (\langle r^2 \rangle_n - \langle r \rangle_n)
    \label{WF}
\end{equation} 

Minimizing this functional yields Maximally Localized Wannier functions (MLWFs) \cite{MLWFs}. MLWFs allow the problem to scale more linearly, and aid in the construction of model Hamiltonians such as those in the TB approximation \cite{tb}:

\begin{equation}
    H(k) = \sum_i E_i(\mathbf{R}_a) c_i^\dag c_i + \sum_{\langle i,j \rangle} t_{ij}(\mathbf{R}_a) c_i^\dag c_j + h.c.
    \label{eq.3}
\end{equation} 

Here $H$ is the TB Hamiltonian, $E_i(R_a)$ is the on-site energy, $t_{ij}(R_a)$ are hopping parameters (between $i$-th orbital in home unit cell at $R=0$ and $j$-th orbital in a unit cell at $\mathbf{R}$), $c_i^\dag$ and $c_i$ are creation and annihilation operators, respectively and $\mathbf{R}_a$ are atomic positions. 

Semi-empirical TB is an intuitive and computationally efficient approach towards evaluating crystal systems, in which the hopping parameters are often obtained from the overlap between the relevant atomic orbitals \cite{tb1}. Accuracy is sacrificed for speed, simplicity, and interpretability. This is in direct contrast to first principles calculations. Wannierized TB models seek a balance, leveraging the simplicity of empirical methods with the accuracy from ab initio calculations. They are important in studying semiclassical theory of electron dynamics that has been implemented in this article. They can yield simplified and highly interpretable insights by significantly reducing the degrees of freedom and providing flexibility in parametrizing the Hamiltonian.

\subsection{\label{tranportproperties}Transport Properties}

Electronic properties of perovskites can be determined through their band structure, obtained by solving the many body Schrodinger equation, and the desnity of states (DOS) given by:

\begin{equation}
    g(E) = \frac{1}{2 \pi^2}\bigg(\frac{2m^*E}{h^2}\bigg)^{3/2} E^{1/2}
    \label{DOS}
\end{equation} 
where, E is the dispersion relation (${h^2k^2}/{2m^*}$).

Thermoelectric transport properties are quantified by electrical ($\sigma$) and thermal ($\kappa$) conductivity, and Seebeck coefficient ($S$) at a specific temperature. Boltzmann transport equation provide the semiclassical treatment to study the transport properties of materials with complex band structures \cite{boltzwann}. 

In semiclassical transport theory framework, the expressions for $\sigma$, $S$ and $\kappa$ tensors are:

\begin{equation}
    \sigma = e^2\int \Sigma_E(E) \frac{-\partial f}{\partial E} dE
    \label{sigma}
\end{equation} 
    
\begin{equation}
    S = \frac{e}{T}\int \Sigma_E(E-E_F) \frac{-\partial f}{\partial E}dE
    \label{seebeck}
\end{equation} 
   
\begin{equation}
     \kappa_{el} = \frac{1}{T} \int \Sigma_E(E-E_F)^2 \frac{-\partial f}{\partial E} - TS^2\sigma
    \label{sigma}
\end{equation} 

where, $E_F$ is the fermi energy, $f$ is the Fermi-Dirac distribution, and $\Sigma_E$ is the transport distribution function (TDF), a tensor sum over $v_i$, the $i$-th component of band velocity, and the relaxation time $\tau$:

\begin{equation}
     \Sigma_{E_{ij}}(E) = \frac{1}{V}\sum_{n,k} v_i(n,k)v_j(n,k)\tau(n,k)\delta(E-E_{n,k})
    \label{sigma}
\end{equation} 

Here, V is the volume of system, $n$ is the band index at crystal momentum, $k$.

These parameters then collectively control the Figure of Merit (ZT), the principal metric used to judge thermoelectric materials:

 \begin{equation}
    ZT = \frac{\sigma S^2 T}{\kappa}
    \label{FOM}
\end{equation} 

High performing thermoelectric materials (high ZT) should have high $\sigma$ and $S$, and low $\kappa$.

\section{\label{methodology}Methodology}

\begin{figure}

    \includegraphics[scale=0.40]{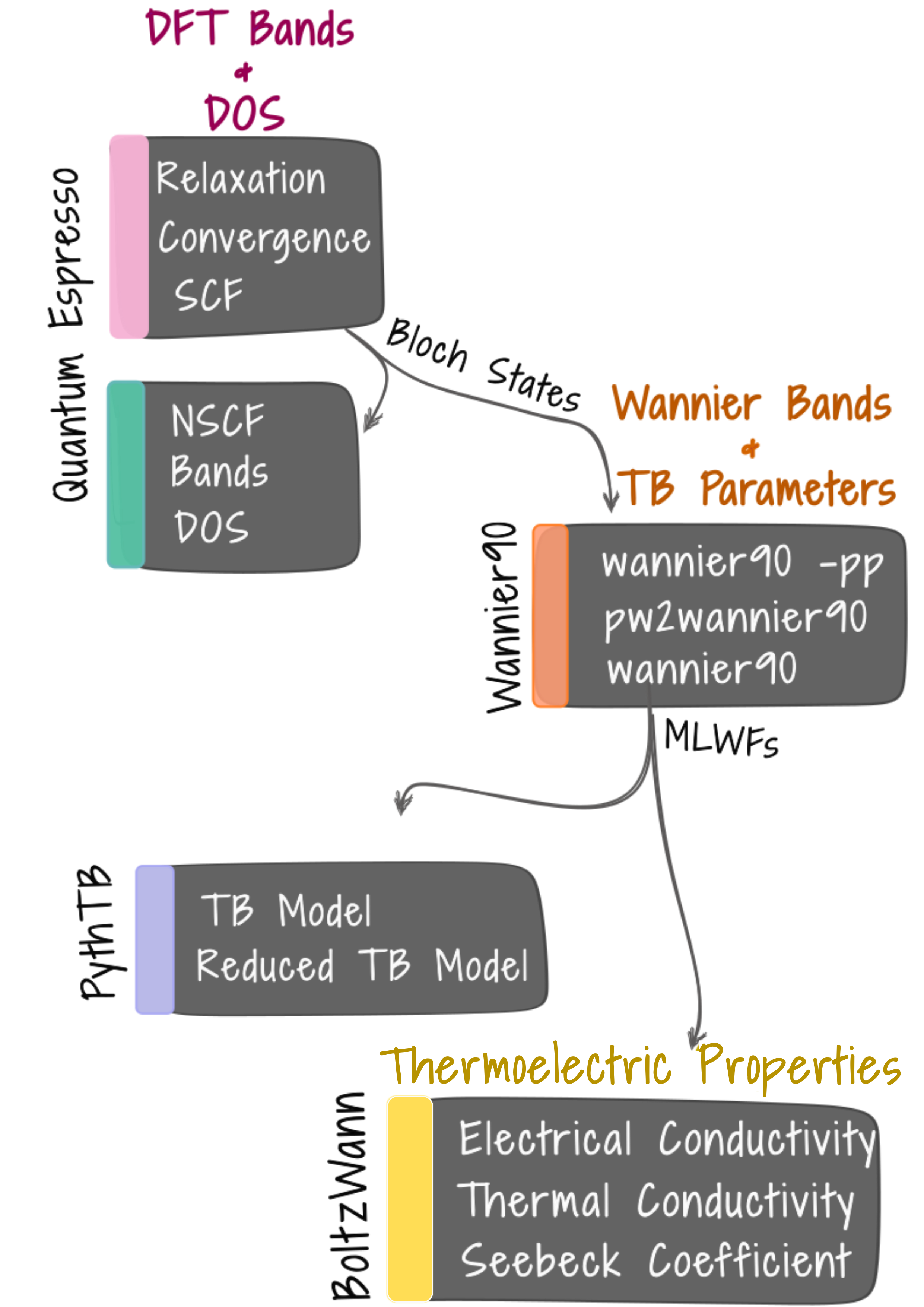}
    \caption{Workflow showing the detailed steps followed for computations. The Bloch states computed using DFT were fed for wannierization. The MLWFs thus generated were utilized for the TB models and thermoelectric evaluation.}
    \label{workflow}
\end{figure}

In order to investigate the inorganic lead halide perovskites with partial and complete anion exchange for their structural, electronic and thermoelectric properties, we followed the four step workflow presented in Fig.~\ref{workflow}. 

\emph{Structure Generation:} The four crystalline perovskite structures, \Briii\, \BriiI\, \BrIii\, and \Iiii\ were generated using Vesta \cite{Momma:db5098} and relaxed using Quantum Espresso. Table ~\ref{structuralinfo} shows the geometrical details of perovskite structures: Goldschmidt tolerance factor ($t$), octahedral factor ($\mu$), unit cell volume, and electronegativity. 
\begin{table*}
\caption{\label{structuralinfo}Structural properties of perovskites}
\begin{ruledtabular}
\begin{tabular}{lrrrr}
 
 Material & Goldschmidt tolerance & Octahedral factor & Unit cell Volume $(a.u)^3$& Electronegativity \\
         \hline
         \Briii     &   0.862   & 0.61  & 1455  & 4.55 \\
         \BriiI     &   0.858   & 0.58  & 1553  & 4.47\\
         \BrIii     &   0.855   &0.56   & 1653  & 4.34 \\
         \Iiii      &   0.851   & 0.54  & 1623  & 4.28\\
\end{tabular}
\end{ruledtabular}
\end{table*}
The structures are cubic and pseudocubic with $Pm\overline{3}m$ space group. The parameters $t$ and $\mu$, were calculated using the equations ~(\ref{eq.1}) and ~(\ref{eq.2}):

\begin{equation}
    t = \frac {r_A + r_X} {\sqrt{2} (r_B + r_X)}
    \label{eq.1}
\end{equation}

\begin{equation}
    \mu = \frac{r_B}{r_X}
    \label{eq.2}
\end{equation} 

where $r_A$, $r_B$ and $r_X$ are the Shannon radii of the Cs, Pb and X=Br and I, respectively in the perovskite structure.

\emph{DFT Calculations:}
DFT calculations were performed using the Generalized Gradient Approximation (GGA) Perdew-Burke-Ernzerhof (PBE) exchange-correlation (XC) with Quantum Espresso (QE), version 7.2 \cite{QE-2009} \cite{doi:10.1063/5.0005082} to obtain the bandstructure and density of states (DOS). Variable cell relaxation was performed for each structure prior to calculations. Convergence tests were carried out for ecutwfc, ecutrho, and Monkhorst-Pack k-points mesh. Further details of convergence parameters are given in Table ~\ref{DFTparams}. 

\emph{Wannierization and Tight-Binding:}
Parametrized tight binding model was created using Wannier90 (W90) \cite{MOSTOFI20142309} and pythtb \cite{coh_vanderbilt_2022}. The pythtb code was modified to generate TB as well as redTB model. A twelve member basis set was selected for p orbitals of Pb and halogen atoms present in perovskite structures. 


Pythtb code was implemeted to extract the on-site energies and hopping parameters for both TB and RedTB models. TB model utilzies all the contributions however RedTB model require bare minimum thresold of nearest neighbors and hopping norm.

\emph{Thermoelectric Properties:}
Finally, the developed tb model has been applied to study thermoelectric properties of perovskites through utilizing the knowledge of transport phenomenon implemented through Boltzmann transport equation. Calculations were performed using Boltzwann module of postw90 by varying the $\mu$ from in the vicinity of fermi energy (as bands only in this region with width of $k_B T$ around fermi energy are relevant). Figure of merit ZT (equation (9)) is calculated after obtaining the elecrtical condcutivity ($\sigma$), Seebeck coefficient ($S$), thermal conductivity ($\kappa = \kappa_{electronic} + \kappa_{Lattice}$) at different temperatures $T$.

\section{\label{resultsanddiscussion}Results and Discussion}

\subsection{\label{dftresults}DFT Results}

\begin{figure*}
    \includegraphics[scale=0.60]{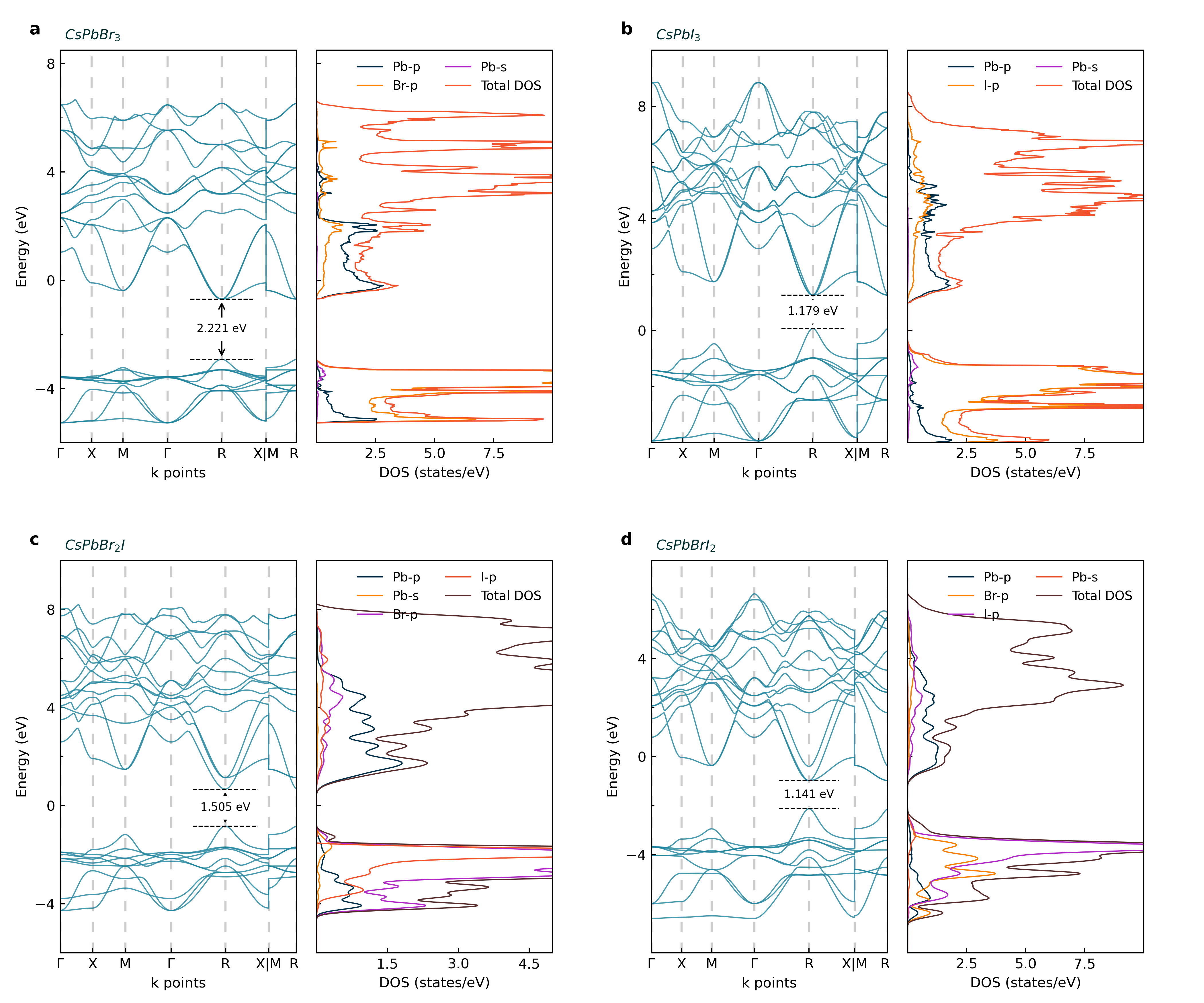}
    \caption{\label{DFTbandDOS}Band structure along high symmetry points and density of states for perovskites obtained from DFT calculations. Anion exchange tuned the bandgaps resulting in maximum bandgap for \Briii\ and minimum bandgap for \BriiI\ .}
\end{figure*}

The DFT generated band structure along the selected k-path and the corresponding density of states (DOS) for the materials, with contributions from the lead and halogen s and p orbitals, are shown in Fig.~\ref{DFTbandDOS}, with all four systems having direct bandgaps at R. As is usual for perovskites, Cs orbitals play more of a stabilizing role and contribute relatively little to the bands or the band gap. Their DOS contributions reside in the higher end of the conduction band, away from the region of interest, and are not shown. 

The valence bands comprise principally the nine halogen centered p orbitals with some hybridization induced contribution from the Pb-s and Pb-p orbitals whereas the Pb-p and X-p orbitals feature prominently in the lower half of the conduction band. \Briii\ and \Iiii\ show a high degree of degeneracy which is lifted as we move to their less symmetric anion exchanged counterparts, \BriiI\ and \BrIii. In addition to its significantly smaller bandgap, \BrIii\ bands also exhibit dense packing and small dispersion attributable to the slight shifting of halogen atom with reference to Pb atom \cite{doi:10.1021/jp5126365}. 

Thermoelectric bandgaps need to be large enough to isolate n-type and p-type carriers, but small enough to assist mobility. 
The 2.221 eV bandgap shown by \Briii\ can be tuned by keeping the A and B cations constant while swapping the X anions, thus giving us bandgaps in the order \Briii $>$ \BriiI $>$ \Iiii $>$ \BrIii. 

\begin{figure}
    \includegraphics[width=3in]{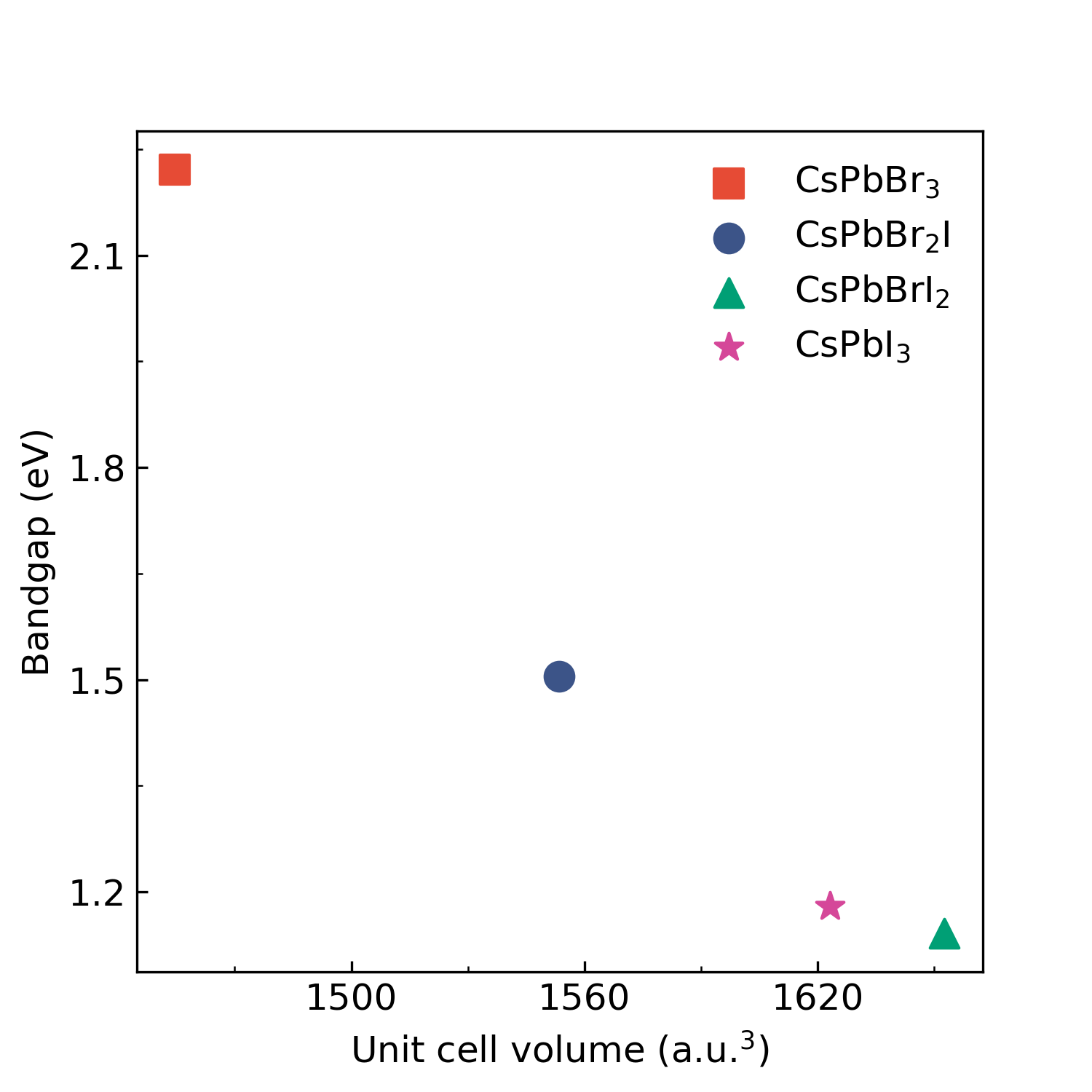}
    \caption{\label{bgvol}Relationship between cell volume and bandgap shows a downward trend because of the reduced orbital overlap.}
\end{figure}

\begin{figure}
    \includegraphics[width = 3in]{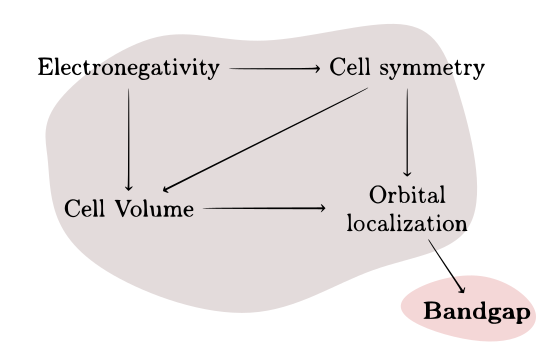}
    \caption{\label{DAG}Relation between different structural and chemical properties that influence the bandgap of perovskites}
\end{figure}

Carrying out partial substitution on the unary halogen \Briii, and \Iiii, to yield the binary halogens \BriiI\ and \BrIii\ results in a loss of symmetry, as evidenced by the lattice parameters, Goldschmidt tolerance $t$ and octahedral tilt $\mu$ in Table ~\ref{structuralinfo}. 
The reduced symmetry increases the overlap between orbitals which narrows the bandgap.
Going from \Briii\ to \Iiii, brings the valence band from 4p to 5p. 
The decreased electronegativity, and hence binding energy and ionization potential, also lead to an increase in the unit cell volume. 
The reduced the orbital overlap between the metal and halogen atoms, thus reduces the bandgap by influencing the edge states of the valence and conduction bands. 
Fig. ~\ref{bgvol} shows this inverse relationship between bandgap and the unit cell volume. 
A slight increase takes place as we restore symmetry in going from \BrIii\ to \Iiii because the volume is optimized to a cubic minima.
The interplay between cell geometry and electronegativity effected through anion exchange ultimately influences the bandgap via the orbital localization and is conceptualized in Fig. ~\ref{DAG}. The combined effect of all these competing factors is mediated through the shifting bandgap \cite{castelli2014bandgap}.

\subsection{\label{wannierandtb}Wannierization and Tight-binding}

\begin{figure}
    \includegraphics[scale=0.4]{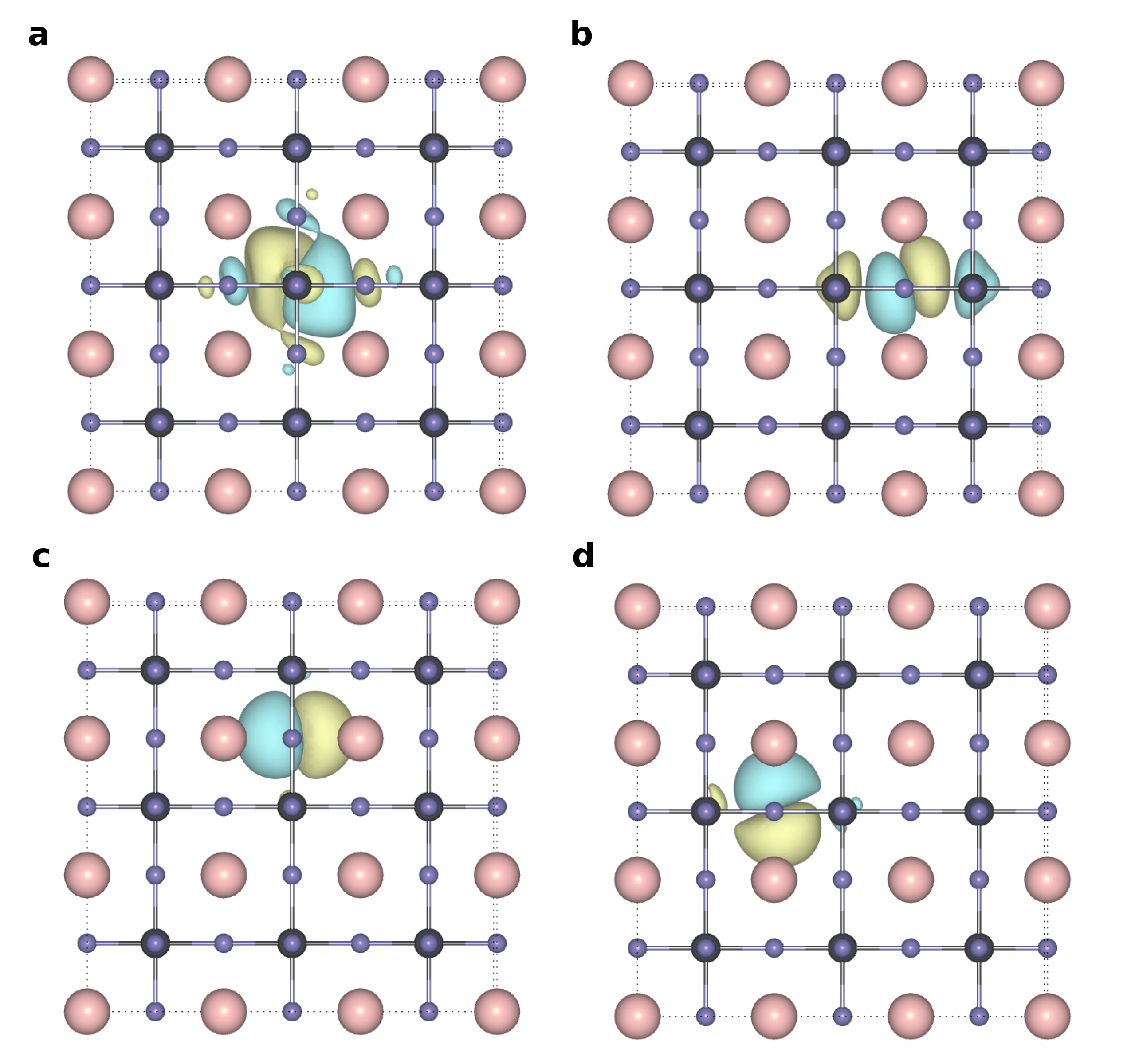}
    \caption{\label{w90orbs}Heat maps showing the tight binding scheme with twelve wannier orbitals. Diagonal entries are on-site energeies and off-diagonal values are hopping parameters with spreads of wannier orbitals are displayed on top of the grid (Pb: dark green, Br: medium green, I: light green). The color bar shows the extent of overlap with green being greater. \BrIii\ shows a uniform field, where as \Iiii\ is more selective.}
\end{figure}

The band structures when regenerated post Wannierization follow the DFT generated profile quite closely ~\ref{w90bands}. The Wannier orbitals for \Briii\ thus obtained are shown in in Fig. ~\ref{w90orbs}. See Fig.  ~\ref{w90orbs-CsPbBr2I}\textendash ~\ref{w90orbs-CsPbI3} for the remaining materials. The Wannierization employed a 12 orbital basis set: the \px, \py, and \pz\ orbitals for one Pb and three X atoms, with one p orbital on each halogen atom directed along the Pb-X-Pb axis. 

\begin{figure*}
    \includegraphics[scale= 0.11]{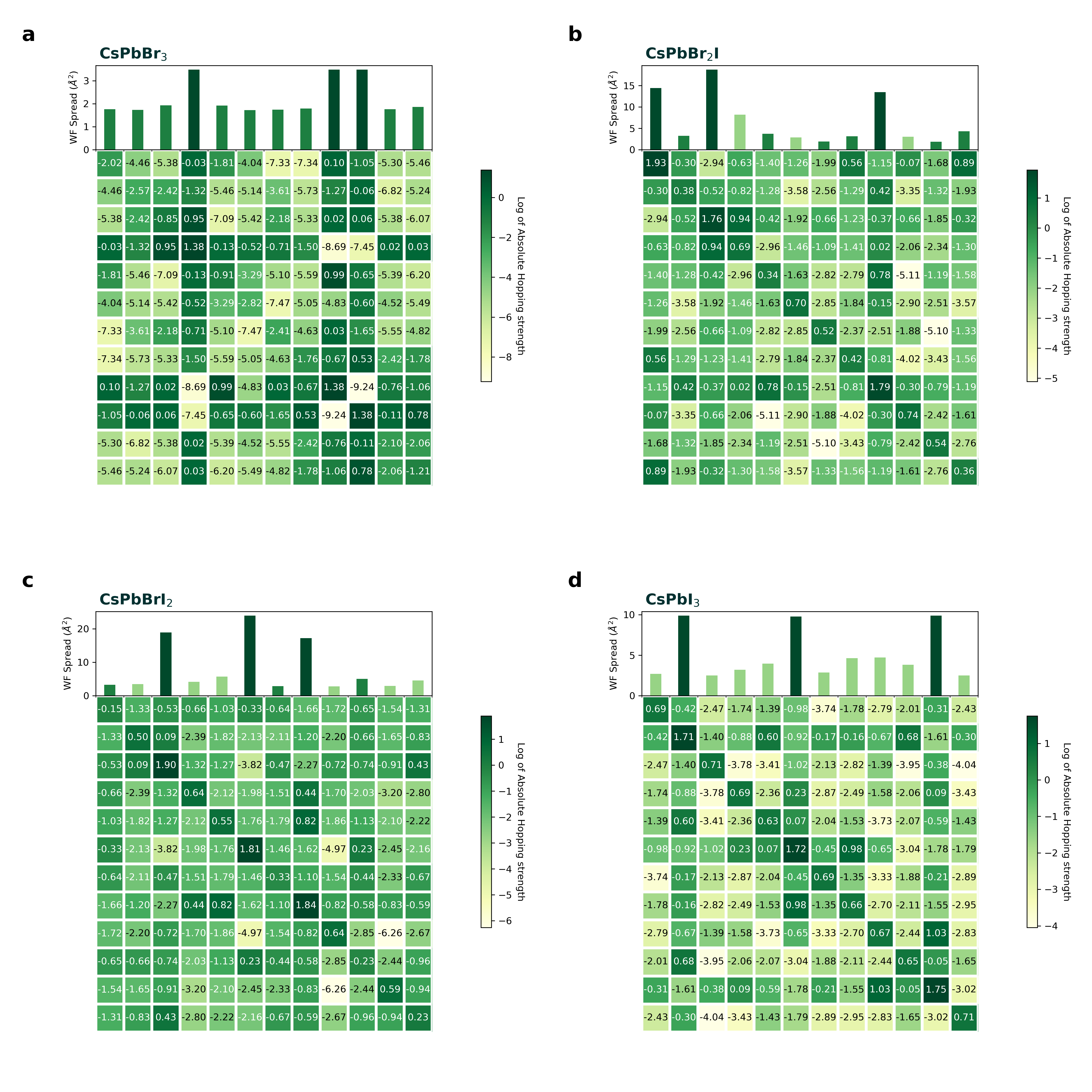}
    \caption{\label{TBgrid}Heat maps showing the tight binding scheme with twelve wannier orbitals. Diagonal entries are on-site energeies and off-diagonal values are hopping parameters with spreads of wannier orbitals are displayed on top of the grid (Pb: dark green, Br: medium green, I: light green). The color bar shows the extent of overlap with green being greater. \BrIii\ shows a uniform field, where as \Iiii\ is more selective.}
\end{figure*}

Wannierization allows the construction of a tight binding (TB) model. The $3\times3\times3$ supercell generated models are shown as heat maps  along with the spread of Wannier functions in Fig. ~\ref{TBgrid}. The log of on-site energies are placed along the diagonal and hopping strengths off it. As a sanity check it can be seen that Pb-Pb hoppings, say for \Briii\ are small due to their orthogonal nature but large between Pb and Br for row 3 because of its orientation along the line Pb-X-Pb. 

It will be noted that the heat maps are more uniform in the case of \BriiI\ and \BrIii\ . 

By comparison, the more delocalized \BriiI\ contains stronger hoppings, from farther away. Put another way, the ionic behavior in \BrIii\ is thus less selective and more gregarious than say \Iiii.
Moreover, the orbitals (coded as dark green for Pb, medium green for Br, and light green for I) already considerably spread out for the Pb centered p orbitals relative to the X, delocalize further for the iodine based materials, for example \BrIii\ versus \Briii. This is again a manifestation of the loss of symmetry. 
In other words, the push towards delocalizaton through increased cell volume and lower electronegativity is countered to some extent through the loss of symmetry. This leads us to the scheme in Fig. ~\ref{DAG} where the unit cell volume and direction of orbitals both influence the orbital localization and hence the bandgap.
In the TB paradigm we discuss below, this would be effected via the changes in the hopping weights. 

\begin{figure*}
    \includegraphics[scale= 0.14]{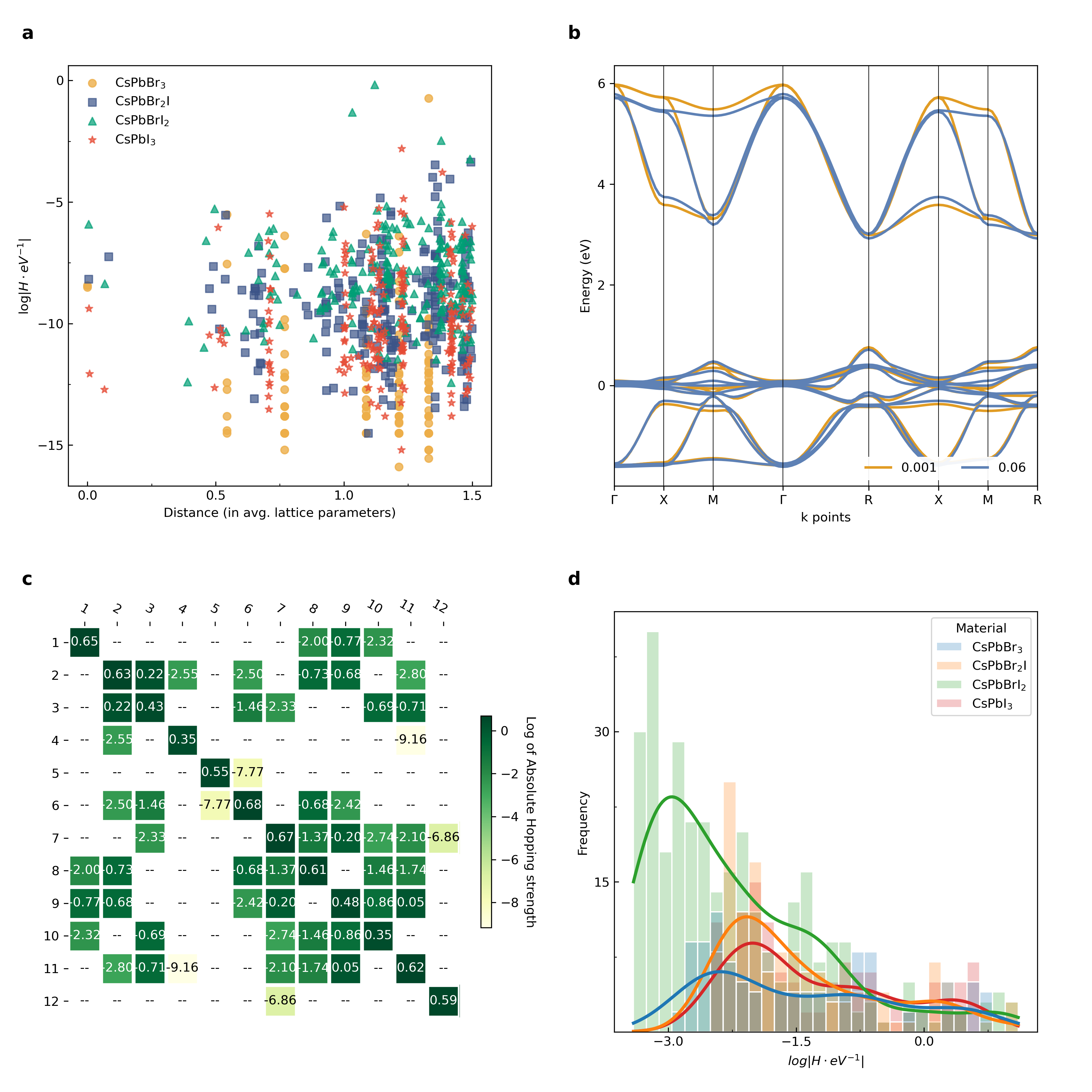}
    \caption{\label{figRedTB} (a) Hopping contributions of all perovskites in terms of distance from home cell. The contributions are localized on specific values of distance for symmetric structures \Briii\ and \Iiii\ , and are scattered over the plot for anion exchanged structures \BriiI\ and \BrIii\ . (b) Wannier bands recovered for \Briii\ in reduced tight binding scheme with minimum hopping norm of 0.06. (c) Heat map for \Briii\ showing the minimum contributions in reduced tight binding scheme that can recover the band structure. (d) Density plot showing the hop distributions for perovskites.}
\end{figure*}

Fig. ~\ref{figRedTB}a shows a randomized 10,000 point sample of the hoppings that lie within 1.5 times the average lattice length.  \Briii\ with its localized orbitals and smaller unit cell has a wider spectrum of hoppings, both weak and strong, within the specified region. Almost similar is the case of \Iiii\ but the delocalized orbitals of \BriiI\ and especially \BrIii\ make more dispersed hopping spectrum over pointing towards the more contributions of weak strength.

The complete TB model contains redundant information. A leaner model, one that uses fewest hops, can be created by adjusting the maximum distance and minimum hopping thresholds. 
The \Briii\ Wannier bands under two thresholds are shown in ~\ref{figRedTB}b. The orange curves have a minimum hopping strength of 0.001 eV and no maximum distance cutoff apart from the dimension of the supercell. On the other hand, with a hopping cutoff of 0.06 eV and a maximum distance cutoff of 1.5 times lattice vector, the blue curve can recovers the key feature of the bands profile with only a fraction of terms (roughly 0.1\%).
This leads to the reduced TB (RedTB) scheme of fig. ~\ref{figRedTB}c which is considerably sparser and shows the most essential hops. Table ~\ref{RedTBparams} contains the threshold value of hopping norm and corresponding number of hopping parameters required to recover the bands. redTB maps along with the recovered bands with minimum hopping norms for other three perovskites are added in figures ~\ref{RedTB_supp_CsPbBr2I}\textendash ~\ref{RedTB_supp_CsPbI3}. As can be seen in Supplementary figure ~\ref{RedTB_supp_CsPbBrI2}, the redTB map for \BrIii\ is unique in maintaining a uniform profile. 
A look at the hop distribution in the reduced framework (Fig. ~\ref{figRedTB}d) gives insight as to why this is so. \Briii\ and \Iiii\ with their confined orbitals, follow a 'few-strong' scheme: a few hops of greater strength are sufficient to capture the essential details of the band structure. The spectrum of \BrIii\ is unique in following a 'many-weak' scheme: many hops, though of weaker strength are needed to recreate the bands profile. 

\begin{figure}[H]
\begin{center}

    \includegraphics[width = 3in]{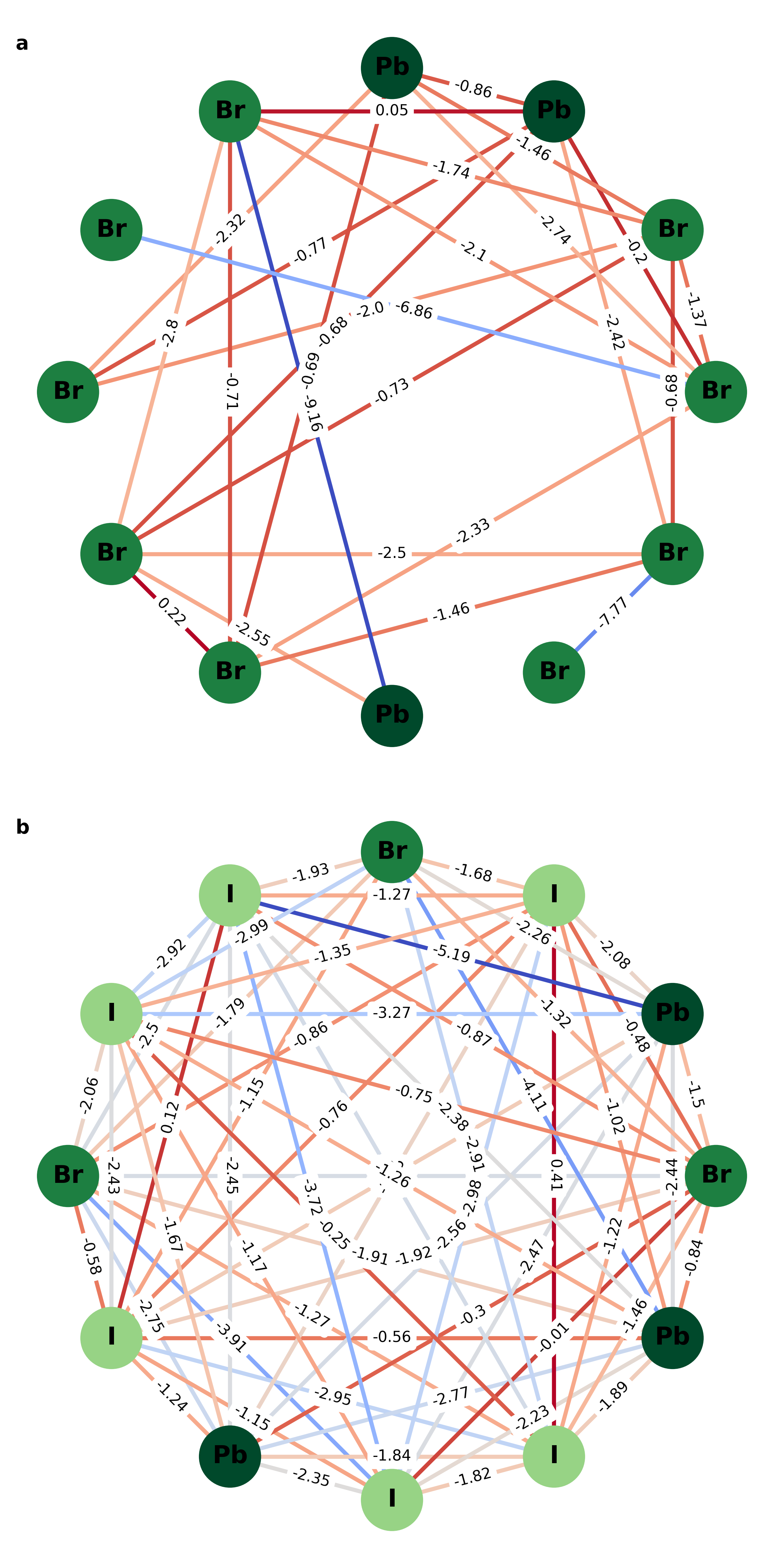}
    \caption{\label{graphgrid}The redTB Hamiltonians expressed as network graphs contrasts the sociable nature of \BrIii\ orbitals \textit{(bottom)} with the reserved behavior of \Briii \textit{(top)}. Connections are colored from red to blue in decreasing values of the log of hopping parameters which are shown on each edge.}
\end{center}
\end{figure}

The 'social' networks of Fig. ~\ref{graphgrid}, expressing the reduced scheme of \Briii\ and \BrIii, show the contrast between the reserved nature of the former with the gregarious nature of the latter: i.e. the few-strong strategy for \Briii\ versus the many-weak one for \BrIii. 

\begin{table*}
\caption{\label{RedTBparams}
The number of parameters in the redTB formalism with the threshold, mean, and standard deviation of log of hopping strengths in eV}
\begin{ruledtabular}
\begin{tabular}{lrrrr}
 
 Material & Num. of parameters & Min. threshold & Mean & Std dev \\
         \hline
         \Briii & 104   &-2.81  & -1.422 & 1.115\\
         \BriiI & 136   &-2.41  & -1.433 & 0.928 \\
         \BrIii &  316  &-3.35  & -2.218 & 1.036 \\
         \Iiii  & 120   &-2.47  & -1.276 & 0.955 \\
\end{tabular}
\end{ruledtabular}
\end{table*}

\subsection{\label{thermoelectricproperties}Thermoelectric Properties}

\begin{figure*}
\includegraphics[scale=0.70]{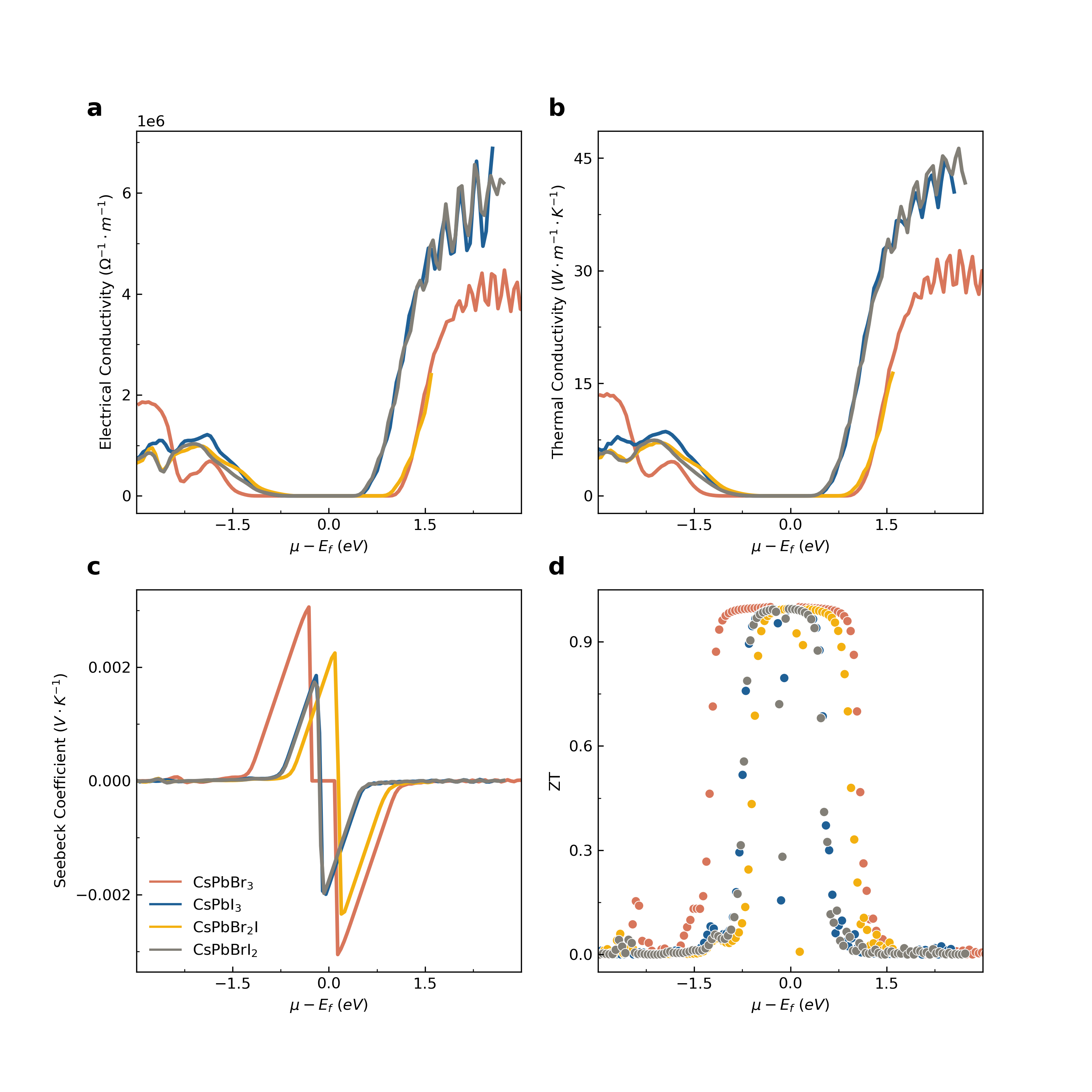}
\caption{\label{figTE}Thermoelectric properties of perovskites showing the (a) Electrical conductivity, (b) Thermal conductivity, (c) Seebeck coefficient and (d) Figure of merit (ZT).}
\end{figure*}

The semiclassical Boltzmann Transport equation describes the electrical transport properties of bulk materials under the influence of an electric field \(\textbf{E}\) and temperature gradient \(\nabla T\). To examine the effect of anion exchange on the thermoelectric properties, the transport properties in relaxation time approximation were evaluated via semi-classical formalism using MLWF basis within the presented TB model \cite{PIZZI2014422}. 

Fig. ~\ref{figTE}(a-d) show the computed thermoelectric properties $\sigma$, $\kappa$, $S$, and ZT vs. the chemical potential ($\mu$) with each curve shifted by its corresponding Fermi energy so that the the axis zero corresponds to the charge neutrality point. 
As mentioned, a material with excellent electrical conductivity, large Seebeck coefficient, and low thermal conductivity would correspond to a high figure of merit. 
Materials with values ZT close to unity in the vicinity of the Fermi level, as we have here, would be regarded as good thermoelectric materials.
Positive (negative) values of the Seebeck coefficient correspond to p-type (n-type) behavior in which holes (electrons) would be the charge carriers. The coefficient, as maybe seen in ~\ref{figTE}(c), is higher close to the charge neutrality point for all four cases with the broadest profile being displayed by \Briii. 
The  larger bandgaps of \Briii\ and \BriiI\ hinders transport and results in lower conductivities. Within the linear region of the plot the electrical and thermal conductivities of \Iiii\ and \BrIii\ are almost three times larger, although these two materials also achieve higher values of the Seebeck coefficient. 
The thermal and conductivities follow a similar profile with a constant ratio. Hence the figure of merit ZT is conditioned predominantly by the Seebeck coefficient. Hence the broad profile of \Briii\ albeit with a low Seebeck coefficient around the Fermi level is reflected in the ~\ref{figTE}(d)
Moreover, the asymmetric nature of the conductivity curves, with higher values for \(\mu > E_f\), indicates that the materials is more suited to electron doping than hole.

\section{\label{conclusion}Conclusion}

Anion exchange allows the tuning of the bandgap to counter the instability of the phases. This begs the question: what is it that happens at the molecular orbital level. Our TB and redTB models generated using MLWFs shed some light on this. The TB profile of \BrIii, which coincidently has the smallest bandgap, is qualitatively and quantitatively different from the other three in the vicinity of the Fermi level, as evidenced by the lack of variation in its TB Hamiltonian. Closer examination via the redTB model and network analysis shows that while the remaining three have a core set of important relationships, the delocalized nature of \BrIii manifests itself in a reduced matrix that shows many hops that are weaker in strength. 

\begin{acknowledgments}
The authors gratefully acknowledge the support of GIK Institute of Engineering Sciences and Technology for providing essential computational facilities for this research. Special appreciation is extended to Dr. Adeel Pervez for providing technical support on various issues.
\end{acknowledgments}

\nocite{*}

\bibliography{references}

\clearpage

\widetext
\setcounter{figure}{0}
\renewcommand{\thefigure}{S\arabic{figure}}
\setcounter{table}{0}
\renewcommand{\thetable}{S\arabic{table}}

\section*{\label{Supplementary Information}Supplementary Information: Density Functional Leveraged Tight-binding Insights into Inorganic Halide Perovskites}
\begin{center}
    Misbah Shaheen and Sheharyar Pervez*\\
Faculty of Engineering Sciences\\
Ghulam Ishaq Khan Institute of Engineering Sciences and Technology
\end{center}

 
         
         
         

\begin{table}[H]
\caption{\label{DFTparams}Input parameters used in DFT calculations.}
\begin{ruledtabular}
\begin{tabular}{lcccccl}
Material & \multicolumn{2}{c}{Threshold (Ry)} & \multicolumn{2}{c}{k-mesh}       & Lattice constants (\AA) \\ 
         & Ecut   & Ecut-rho & SCF             & NSCF             &                          \\ \cline{2-3} \cline{4-5}
\Briii   & 80     & 144      & $8\times8\times8$ & $12\times12\times12$ & $a,b,c = 5.995863$       \\
\BriiI   & 40     & 160      & $8\times8\times8$ & $12\times12\times12$ & $a,b = 5.994375$, $c=6.763426$ \\
\BrIii   & 40     & 160      & $8\times8\times8$ & $12\times12\times12$ & $a,b = 6.402284$, $c=5.659454$ \\
\Iiii    & 40     & 240      & $8\times8\times8$ & $16\times16\times16$ & $a,b,c = 6.219181$       \\
\end{tabular}
\end{ruledtabular}
\end{table}

\clearpage
 \subsection*{\label{w90bands}Wannier Bands and orbitals}

\begin{figure}[H]
    \centering
    \includegraphics[width=7in]{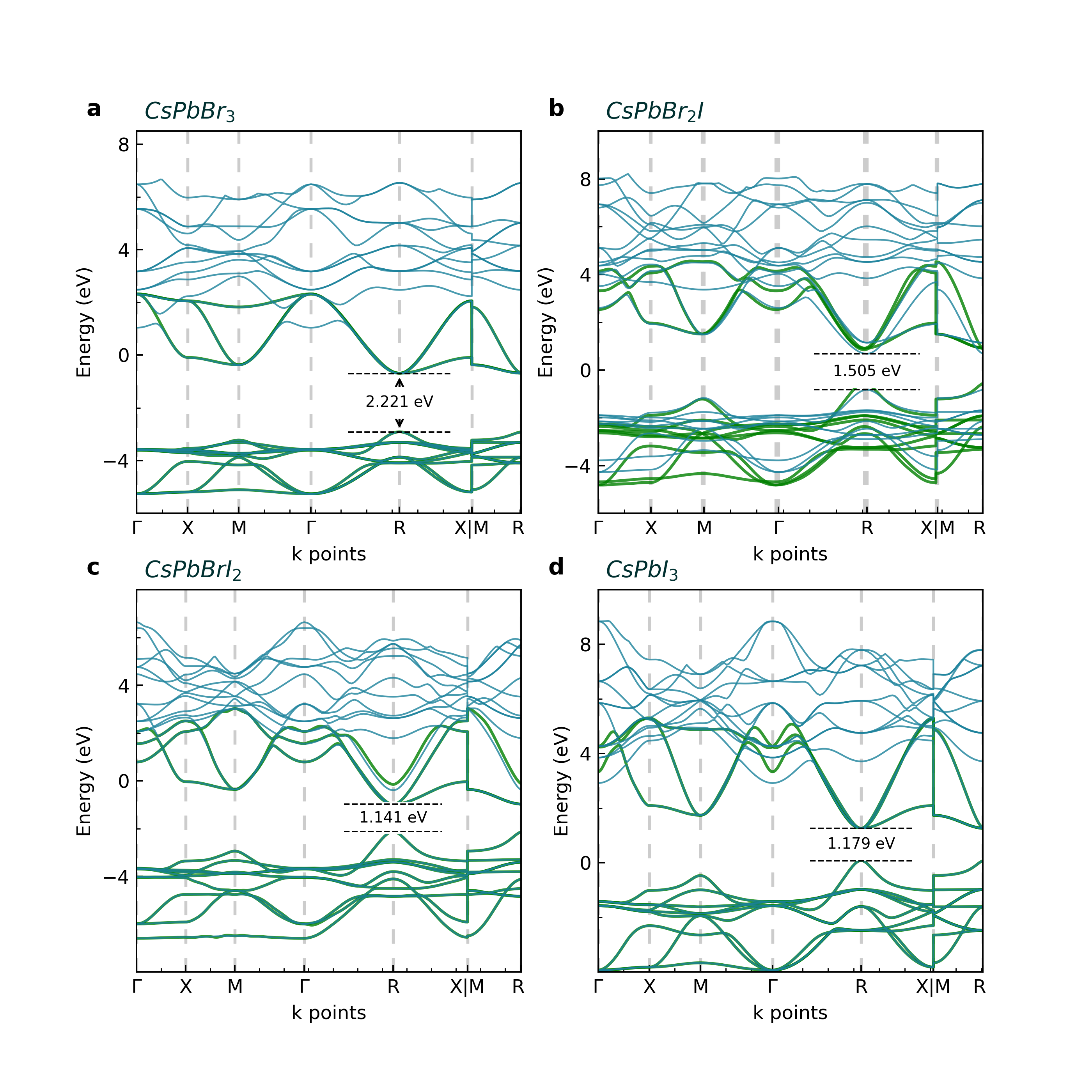}
    \caption{\label{w90bands}Band structure  obtained from Wannierization (green) and from DFT (blue). Twelve Wannier basis were selected for Pb and halogen p-orbitals for calculation.}
\end{figure}

\begin{figure}[H]
\centering
    \begin{subfigure}[c] {0.35\textwidth}
        
        \includegraphics[width=2.2in]{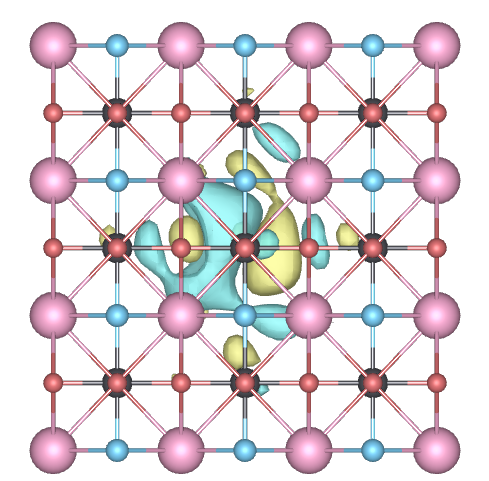} 
        \caption{}
    \end{subfigure}
    \hspace{0.5cm}
    \begin{subfigure}[c]{0.35\textwidth}
        
        \includegraphics[width=2in]{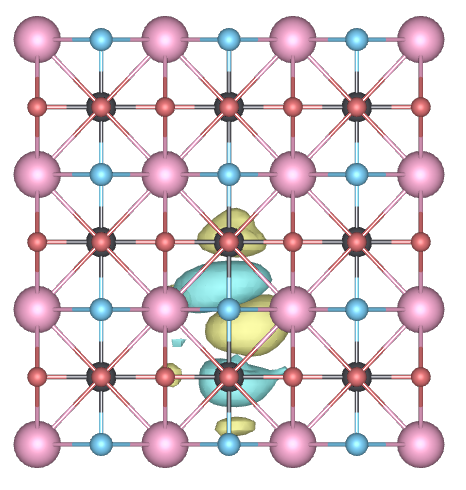} 
        \caption{}
    \end{subfigure}

    \vspace{1em}

    \begin{subfigure}[c]{0.35\textwidth}
        
        \includegraphics[width=2.2in]{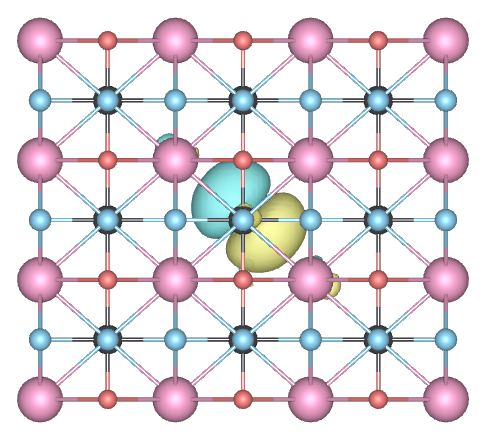} 
        \caption{}
      
    \end{subfigure}
    \hspace{0.5cm}
    \begin{subfigure}[c]{0.35\textwidth}
       
        \includegraphics[width=2in]{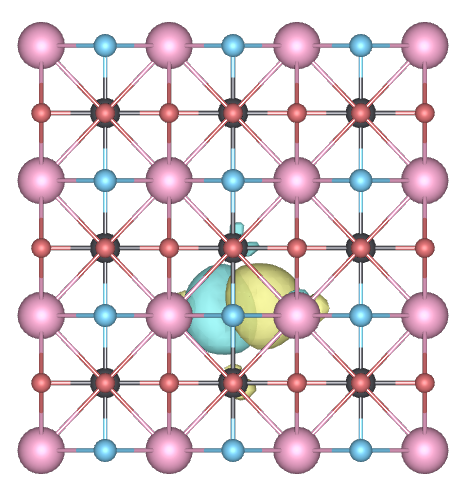} 
        \caption{}
      
    \end{subfigure}

    \caption{Wannier orbitals (yellow and blue isosurfaces) of (a) Pb, (b-d) Halogen for anion exchanged \BriiI\ (Cs: pink, Pb: dark gray, Br: orange and I: blue) perovskite showing delocalized Pb and one of the halogen orbitals along with two relatively localized halogen orbitals. }
    \label{w90orbs-CsPbBr2I}
\end{figure}

\begin{figure}[H]
    \centering
    \begin{subfigure}[c] {0.35\textwidth}
        
        \includegraphics[width=2in]{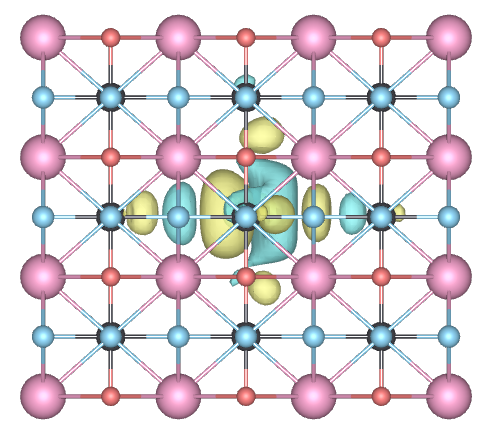} 
        \caption{}
       
    \end{subfigure}
    \hspace{0.5cm}
    \begin{subfigure}[c]{0.35\textwidth}
        
        \includegraphics[width=2in]{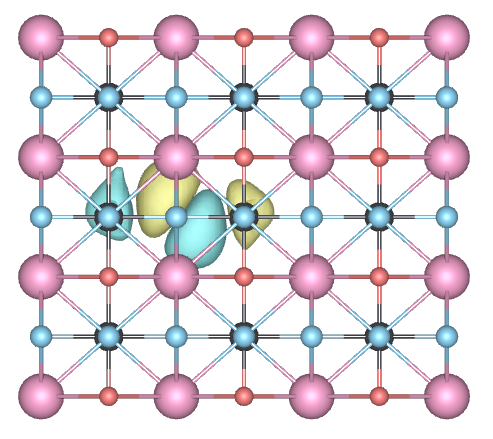} 
        \caption{}
       
    \end{subfigure}

    \vspace{1em}

    \begin{subfigure}[c]{0.35\textwidth}
        
        \includegraphics[width=2in]{_figs/supplementary/w90-orbs-CsPbBrI2/p2.png} 
        \caption{}
       
    \end{subfigure}
    \hspace{0.5cm}
    \begin{subfigure}[c]{0.35\textwidth}
       
        \includegraphics[width=2in]{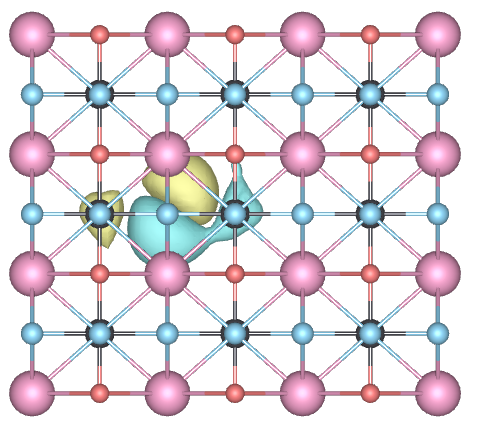} 
        \caption{}
        
    \end{subfigure}

    \caption{Wannier orbitals (yellow and blue isosurfaces) of (a) Pb, (b-d) Halogen for anion exchanged \BrIii\ (Cs: pink, Pb: dark gray, Br: orange and I: blue) perovskite showing delocalized Pb and one of the halogen orbitals along with two relatively localized halogen orbitals. }
    \label{w90orbs-CsPbBrI2}
\end{figure}

\begin{figure}[H]
    \centering
    \begin{subfigure}[c] {0.35\textwidth}
        
        \includegraphics[width=2in]{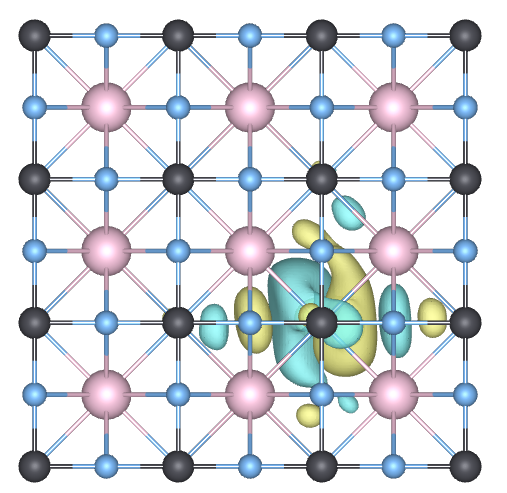} 
        \caption{}

    \end{subfigure}
    \hspace{0.5cm}
    \begin{subfigure}[c]{0.35\textwidth}
        
        \includegraphics[width=2in]{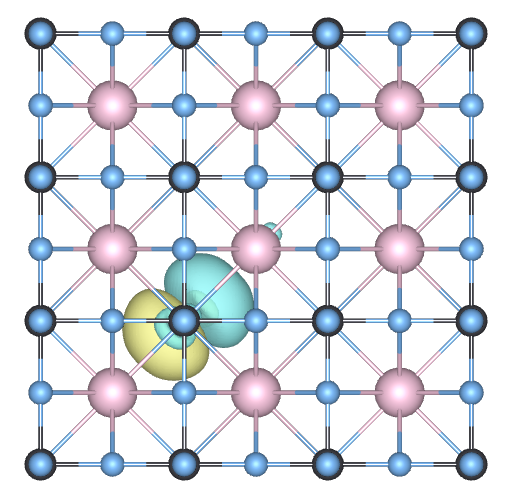} 
        \caption{}
        
    \end{subfigure}

    \vspace{1em}

    \begin{subfigure}[c]{0.35\textwidth}
        
        \includegraphics[width=2in]{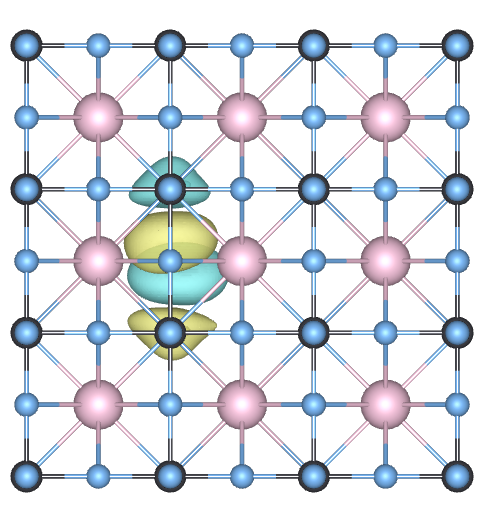} 
        \caption{}
        
    \end{subfigure}
    \hspace{0.5cm}
    \begin{subfigure}[c]{0.35\textwidth}
       
        \includegraphics[width=2in]{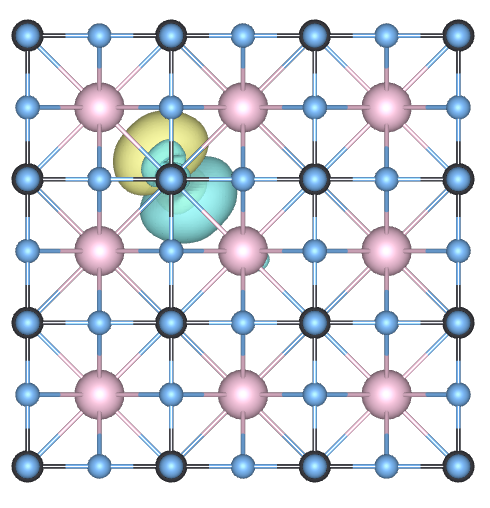} 
        \caption{}
        
    \end{subfigure}

    \caption{Wannier orbitals (yellow and blue isosurfaces) of (a) Pb, (b-d) Halogen for anion exchanged \Iiii\ (Cs: pink, Pb: dark gray, Br: orange and I: blue) perovskite showing delocalized Pb and one of the halogen orbitals along with two relatively localized halogen orbitals. }
    \label{w90orbs-CsPbI3}
\end{figure}

\clearpage
\subsection*{\label{RedTB}Reduced Tight Binding Scheme}

\begin{figure*}[h]
    \centering

    \begin{subfigure}{0.45\textwidth}
        \includegraphics[width=\linewidth]{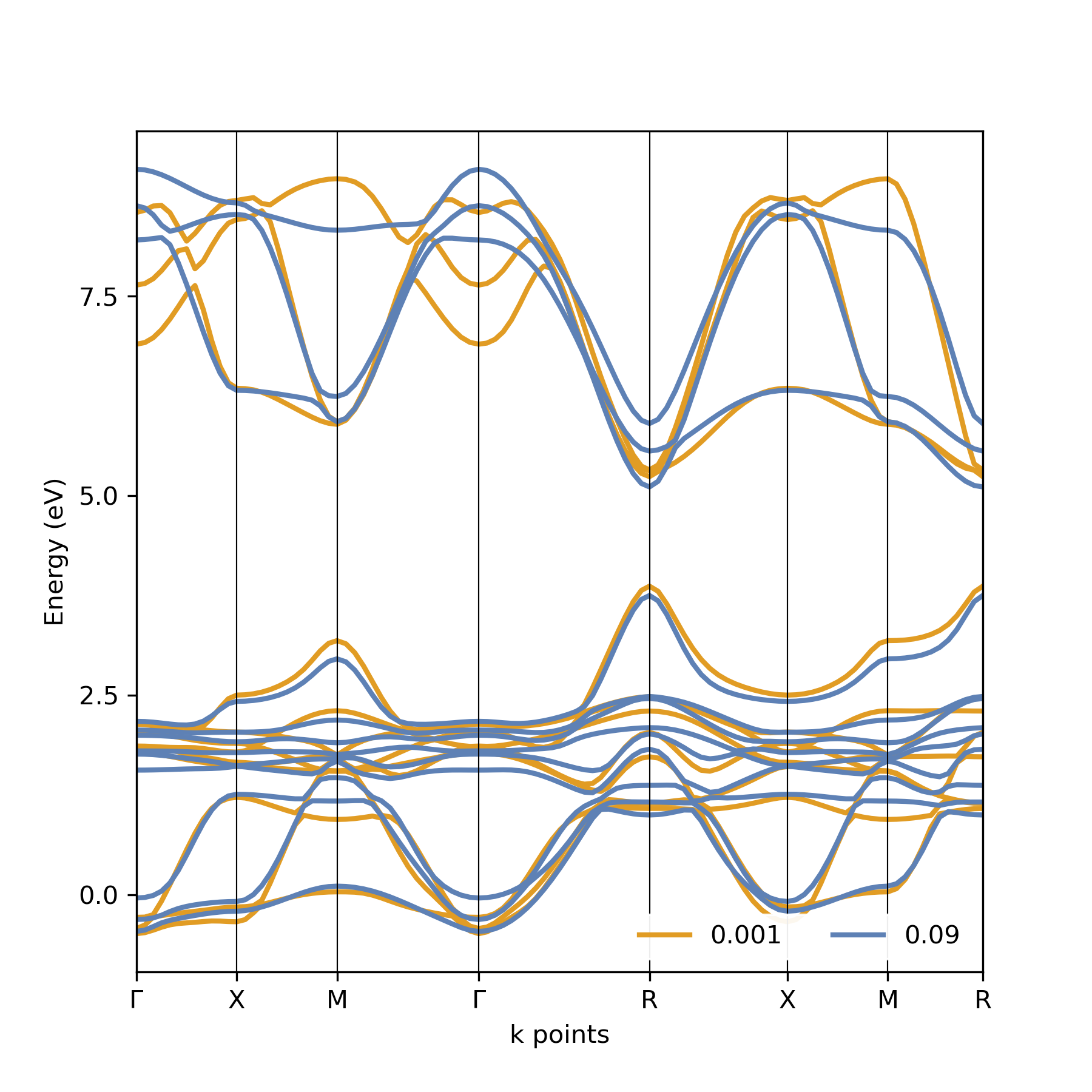}
        \caption{}
    \end{subfigure}
    \hfill
    \begin{subfigure}{0.45\textwidth}
        \includegraphics[width=\linewidth]{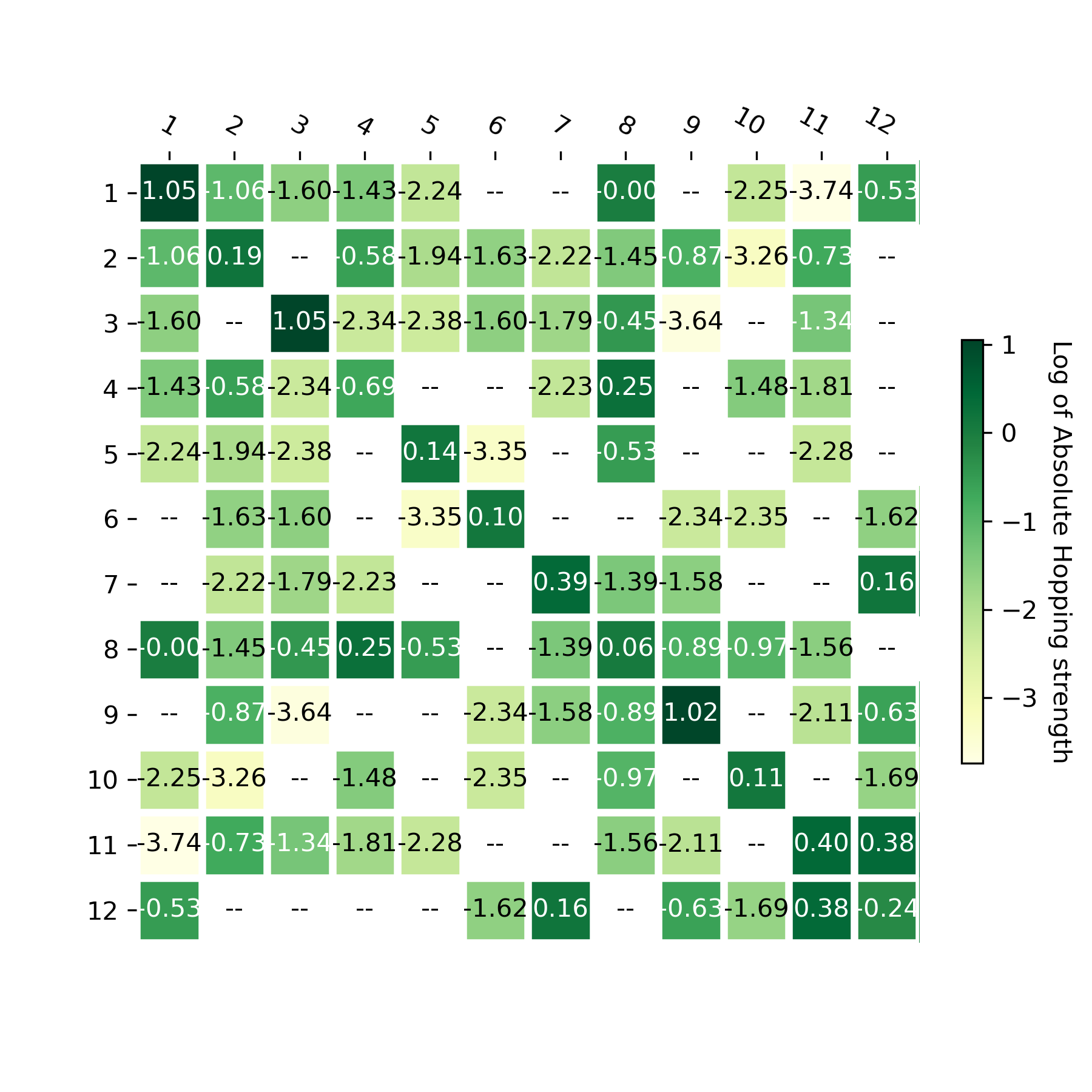}
        \caption{}
    \end{subfigure}
    \caption{(a) Band structure recovered using reduced tight binding scheme with minimum hopping norm of 0.09 (b) Heat map of reduced tight binding scheme for \BriiI\ . }
    \label{RedTB_supp_CsPbBr2I}
\end{figure*}

\begin{figure*}

    \begin{subfigure}{0.45\textwidth}
        \includegraphics[width=\linewidth]{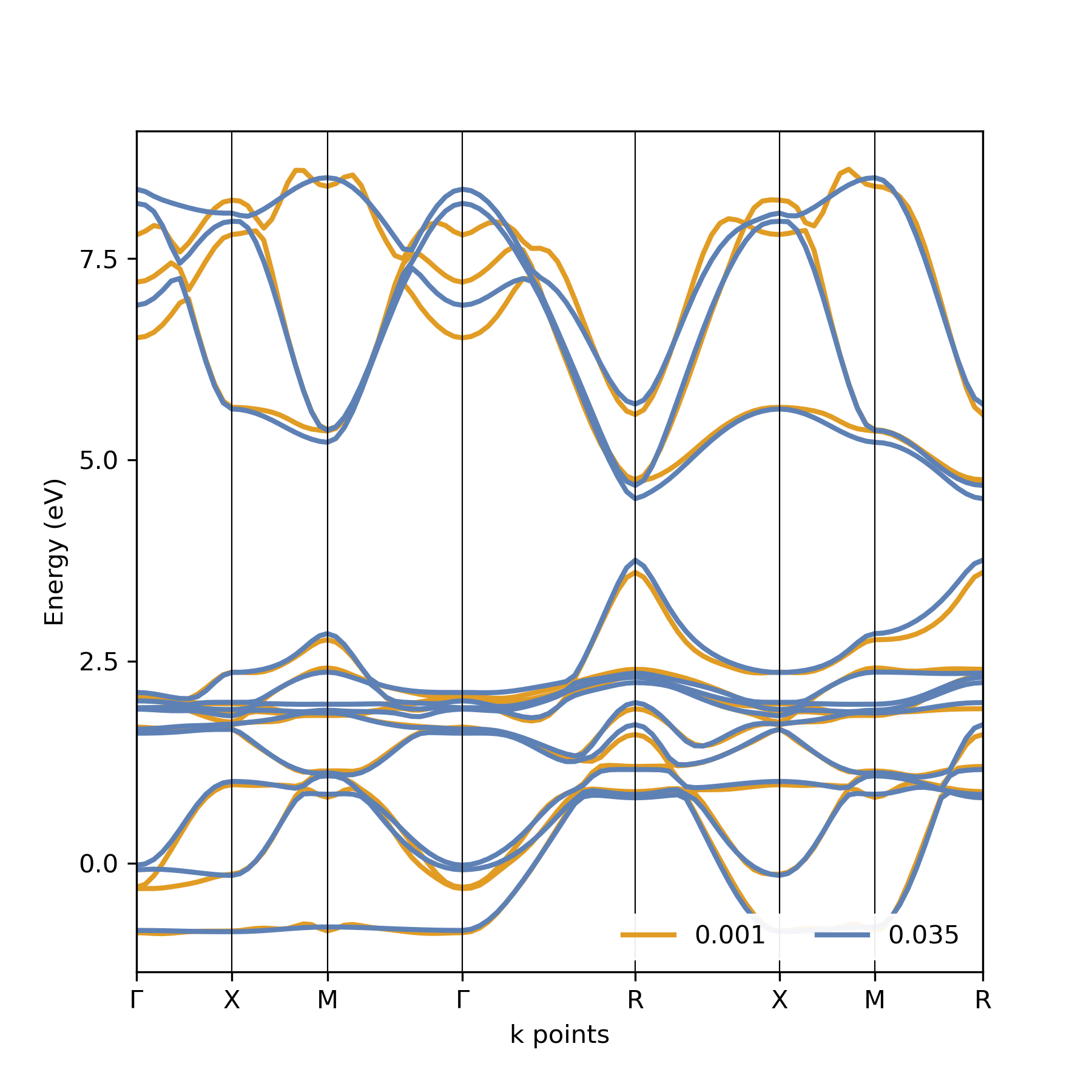}
        \caption{}
    \end{subfigure}
        \begin{subfigure}{0.45\textwidth}
        \includegraphics[width=\linewidth]{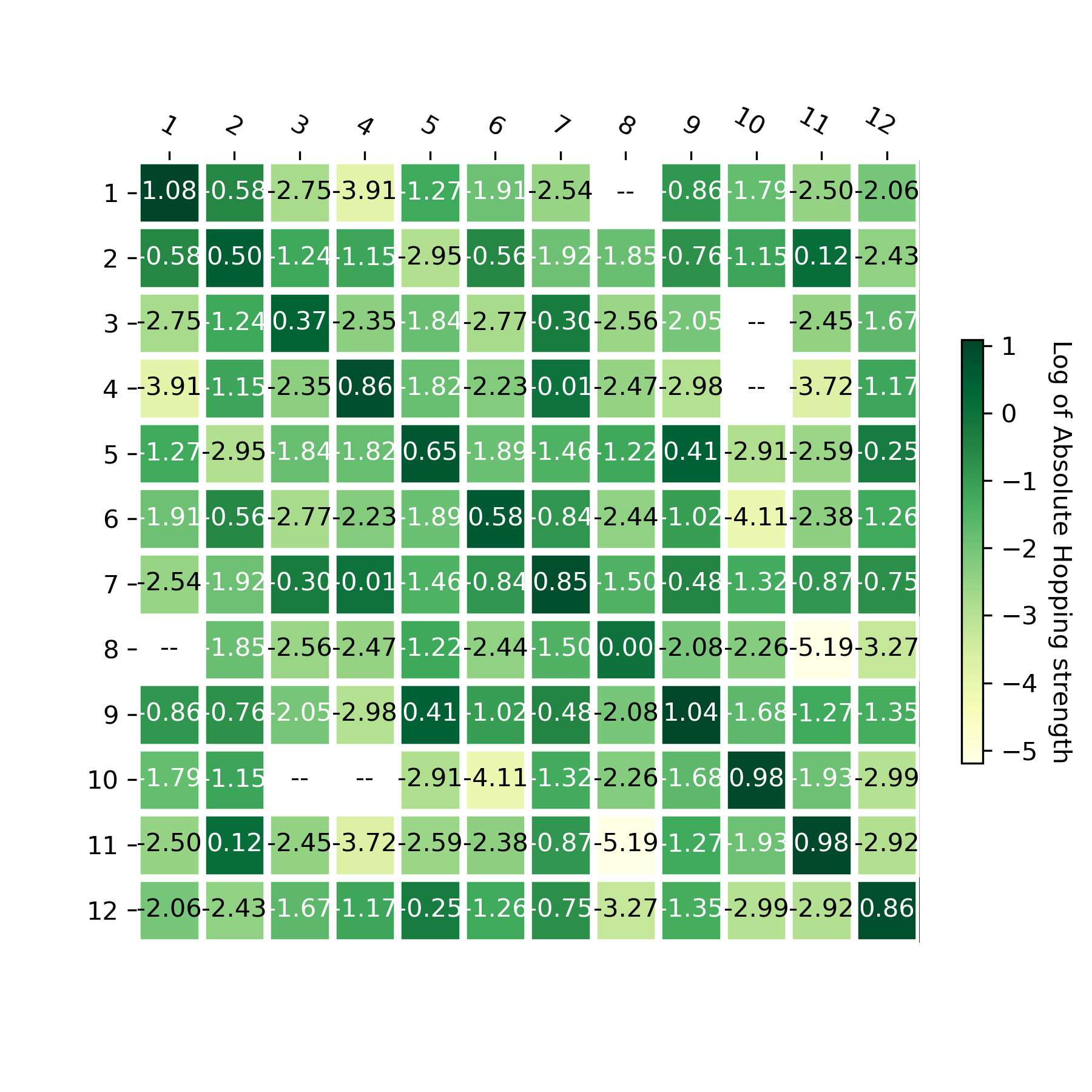}
        \caption{}
    \end{subfigure}

    \caption{(a) Band structure recovered using reduced tight binding scheme with minimum hopping norm of 0.035 (b) Heat map of reduced tight binding scheme for \BrIii\ . }
    \label{RedTB_supp_CsPbBrI2}
\end{figure*}

\begin{figure*}

    \begin{subfigure}{0.45\textwidth}
        \includegraphics[width=\linewidth]{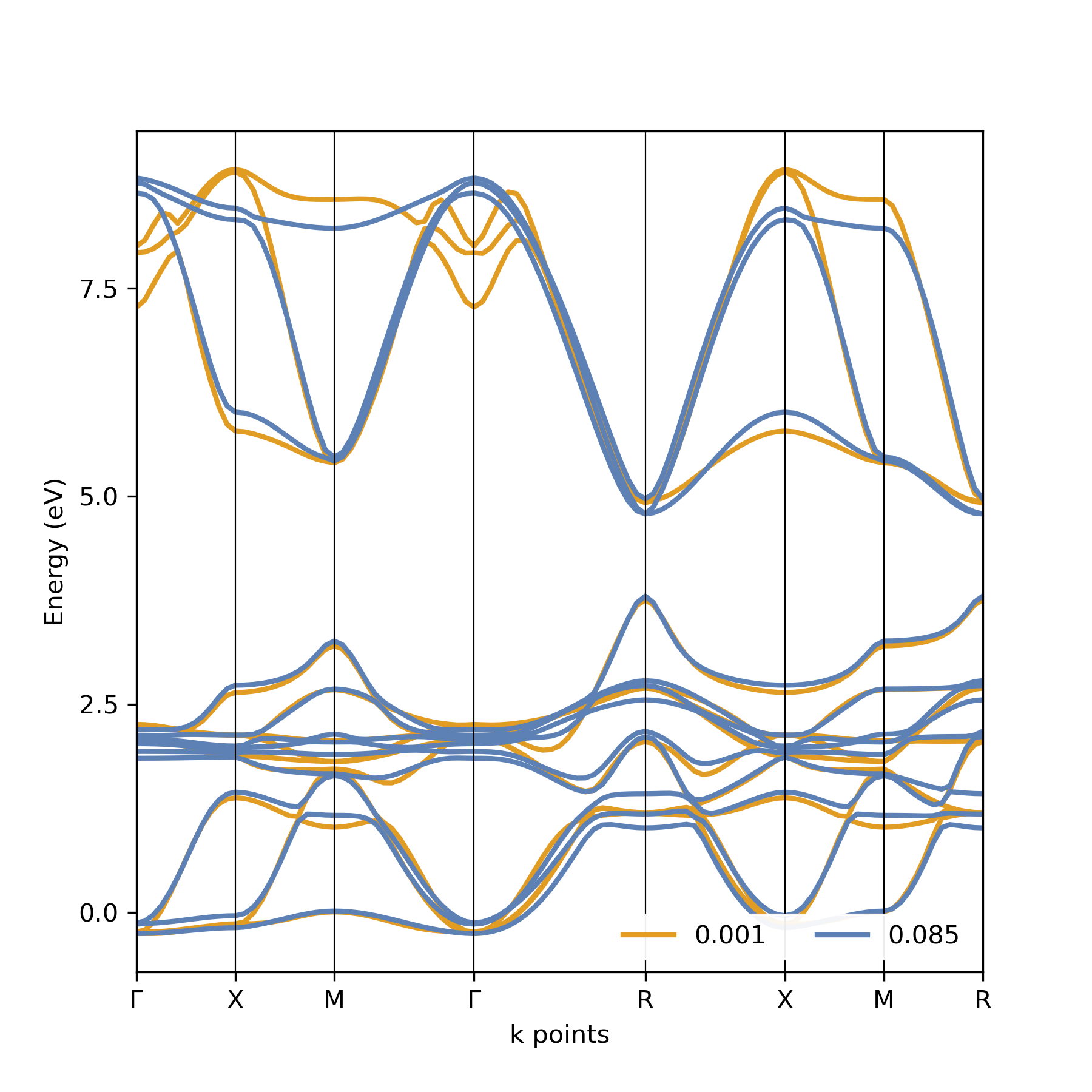}
        \caption{}
        \label{fig:fig3}
    \end{subfigure}
        \begin{subfigure}{0.45\textwidth}
        \includegraphics[width=\linewidth]{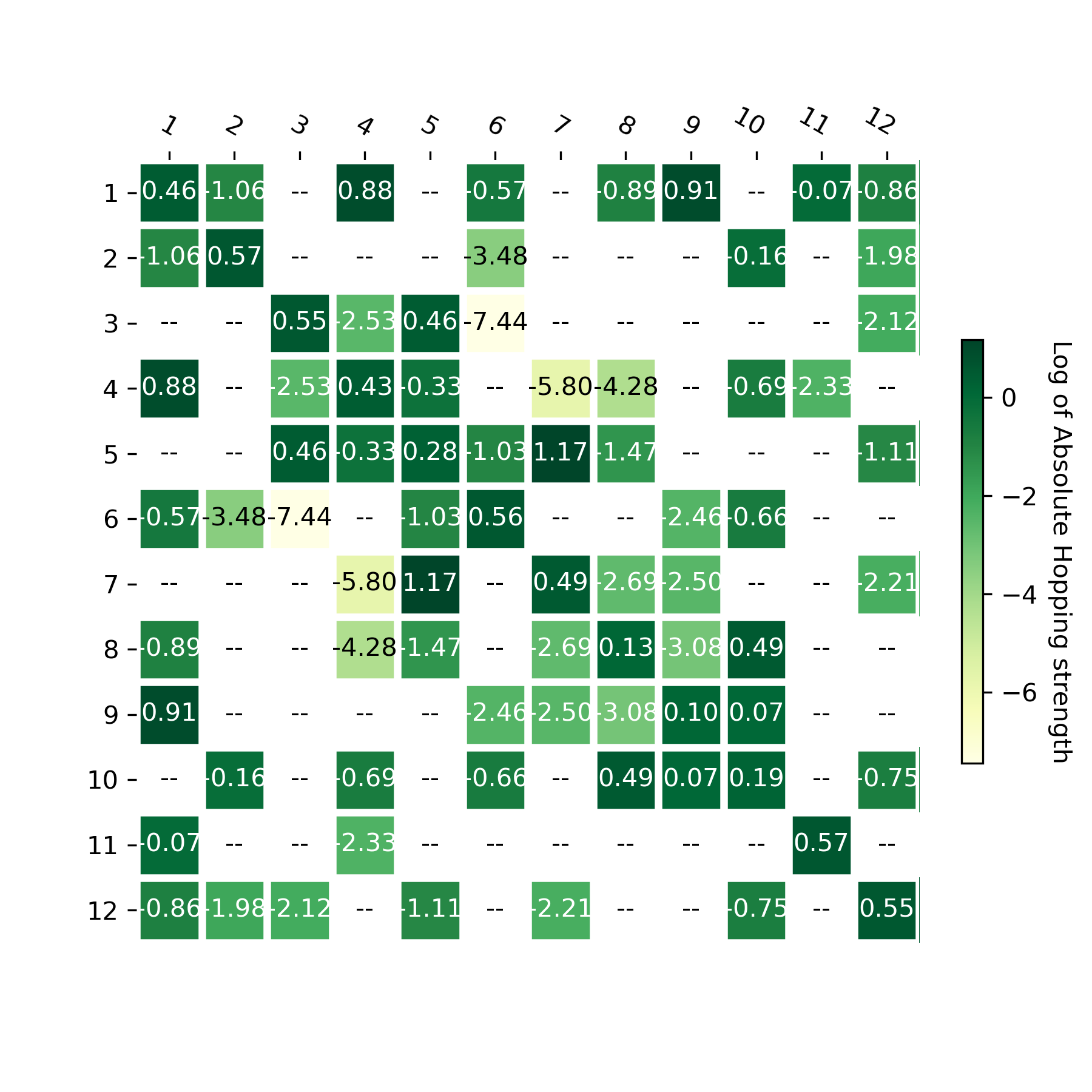}
        \caption{}
        \label{fig:fig3}
    \end{subfigure}

    \caption{(a) Band structure recovered using reduced tight binding scheme with minimum hopping norm of 0.085 (b) Heat map of reduced tight binding scheme for \Iiii\ . }
    \label{RedTB_supp_CsPbI3}
\end{figure*}

\end{document}